\documentclass{iaus}
\usepackage{graphics,epsfig}

  \checkfont{eurm10}
  \iffontfound
    \IfFileExists{upmath.sty}
      {\typeout{^^JFound AMS Euler Roman fonts on the system,
                   using the 'upmath' package.^^J}%
       \usepackage{upmath}}
      {\typeout{^^JFound AMS Euler Roman fonts on the system, but you
                   dont seem to have the}%
       \typeout{'upmath' package installed. iaus.cls can take advantage
                 of these fonts,^^Jif you use 'upmath' package.^^J}%
      }
  \else
  \fi

  \checkfont{msam10}
  \iffontfound
    \IfFileExists{amssymb.sty}
      {\typeout{^^JFound AMS Symbol fonts on the system, using the
                'amssymb' package.^^J}%
       \usepackage{amssymb}%

      }{}
  \fi

  \IfFileExists{amsbsy.sty}
    {\typeout{^^JFound the 'amsbsy' package on the system, using it.^^J}%
     \usepackage{amsbsy}}
    {}

\newsavebox{\astrutbox}
\sbox{\astrutbox}{\rule[-5pt]{0pt}{20pt}}

\title[Stellar Populations in Normal Galaxies]
      {Stellar Populations in Normal Galaxies}

\author[G. Bruzual]{Gustavo Bruzual A.$^1$}

\affiliation{$^1$CIDA, Apartado Postal 264, M\'erida, Venezuela,
email: bruzual@cida.ve}

\pubyear{2004}
\volume{222}
\pagerange{1--8}
\date{?? and in revised form ??}
\jname{The Interplay among Black Holes, Stars and ISM \\in Galactic Nuclei}
\editors{Th. Storchi Bergmann, L.C. Ho \& H.R. Schmitt, eds.}
\begin{document}

\maketitle

\begin{abstract}
I describe very briefly the new libraries of empirical spectra of stars
covering wide ranges of values of the atmospheric parameters T$_{eff}$,
log g, [Fe/H], as well as spectral type, that have become available in the
recent past, among them the HNGSL, MILES, UVES-POP, and Indo-US libraries.
I show the results of using the HNGSL to build population synthesis models.
These libraries are complementary in spectral resolution and wavelength coverage,
and will prove extremely useful to describe spectral features
expected in galaxy spectra from the NUV to the NIR.
\end{abstract}

\section{Introduction}
In a recent paper \cite{BC03}, hereafter BC03, have examined in detail the
advantages of using intermediate resolution stellar spectra in population
synthesis and galaxy evolution models. In these models BC03 use the STELIB
library compiled by \cite{STELIB03}.
This library contains observed spectra of 249 stars in a wide range of
metallicities in the wavelength range from 3200 \AA\ to 9500 \AA\ at a
resolution of 3 \AA\ FWHM (corresponding to a median resolving power of
$\lambda / \Delta\lambda \approx 2000$), with a sampling interval of 1 \AA\ 
and a signal-to-noise ratio of typically 50 per pixel.
The BC03 models reproduce in detail typical galaxy spectra extracted from the
SDSS Early Data Release (\cite{EDR02}).
From these spectral fits one can constrain physical parameters such as the
star formation history, metallicity, and dust content of galaxies
(\cite{HPJD04}, \cite{MAT04}. See also several papers in this conference,
e.g. Cid-Fern\'andez et al.). 
The medium resolution BC03 models enable accurate studies of absorption-line
strengths in galaxies containing stars over the full range of ages and can
reproduce simultaneously the observed strengths of those Lick indices that
do not depend strongly on elemental abundance ratios, provided that the
observed velocity dispersion of the galaxies is accounted for properly,
and offer the possibility to explore new indices over the full optical range
of the STELIB atlas, i.e. 3200 \AA\ to 9500 \AA.
To extend the spectral coverage in the models beyond these limits, we must
recur to other libraries.
For solar metallicity models, the \cite{PIC98} library can be used to extend
the STELIB spectral coverage down to 1150 \AA\ in the UV end and up to 2.5
$\mu$m at the red end, with a sampling interval of 5 \AA\ pixel$^{-1}$ and a
median resolving power of ($\lambda / \Delta\lambda \approx 500$).
The UV spectra in the \cite{PIC98} atlas are based on $IUE$ spectra of bright
stars.
At all metallicities the BC03 models are extended
down to 91 \AA\ in the UV side and up to 160
$\mu$m at the other end using the theoretical model atmospheres included
in the BaSeL series of libraries compiled by \cite{LEJ97}, \cite{LEJ98}, and
\cite{WES02}, but at a resolving power considerably lower than for STELIB
($\lambda / \Delta\lambda \approx 200-500$).
Given the success in reproducing observed galaxy spectra with the synthesis
models built with STELIB, it is important to improve:
first, the spectral resolution outside the STELIB range, and
second, the coverage of the HRD by including more spectral types than those
available in STELIB. These goals are possible now thanks to several
compilations of stellar spectra that have become available in the recent past.
In this paper I describe briefly these libraries and show examples of model
galaxy spectra built using one of these libraries.
The full implementation of the new libraries in the population synthesis models
is in preparation by Bruzual \& Charlot.

\section{New libraries}\label{sec:newlibs}
A large number (4!) of libraries containing medium to high spectral resolution
observed spectra of excellent
quality for large numbers (hundreds to thousand!) of stars have become available
during the last few months. One of the main objectives of the observers
who invested large amounts of time and effort assembling these data sets is
to build libraries suitable for population synthesis. In this respect
the stars in the libraries have been selected to provide broad coverage of
the atmospheric parameters T$_{eff}$, log g, [Fe/H], as well as spectral type,
throughout the HRD. Below I describe very briefly the main
characteristics of each of these libraries, especially those which are
relevant for population synthesis models.

\subsection{HNGSL}
The Hubble's New Generation Spectral Library (\cite{HL03}) contains spectra
for over 200 stars (this number will increase in the future) whose fundamental
parameters, including chemical abundance, are well known from careful analysis
of the visual spectrum. The spectra cover fully the wavelength range from 1700
\AA\ to 10,200 \AA. The advantage of this library over the ones listed below
is the excellent coverage of the near-UV and the range from 9000 \AA\ to 10,200
\AA, which is generally noisy or absent in the other data sets.

\subsection{MILES}
The Medium resolution INT Library of Empirical Spectra (\cite{MILES03}),
contains carefully calibrated and homogeneous quality spectra for 1003 stars
in the wavelength range 3500 \AA\ to 7500 \AA\ with 2 \AA\ spectral resolution
and dispersion 0.9 \AA\ pixel$^{-1}$. The stars included in this library were
chosen aiming at sampling stellar atmospheric parameters as completely as
possible.

\subsection{UVES POP Library}
The UVES Paranal Observatory Project (\cite{UVES03}), has produced a library
of high resolution ($\lambda / \Delta\lambda \approx 80,000$) and high
signal-to-noise ratio spectra for over 400 stars distributed throughout
the HRD. For most of the spectra, the typical final SNR obtained in the V band
is between 300 and 500. The UVES POP library is the richest available database
of observed optical spectral lines.

\subsection{Indo-US library}
The Indo-US library (\cite{VAL04}) contains complete spectra over the entire
3460 \AA\ to 9464 \AA\ wavelength region for 885 stars obtained with the
0.9m Coud\'e Feed telescope at KPNO. The spectral resolution is $\approx$
1 \AA\ and the dispersion 0.44 \AA\ pixel$^{-1}$. The library includes
data for an additional 388 stars, but only with partial spectral coverage.

\subsection{High-spectral resolution theoretical libraries}
There are several on-going efforts to improve the existing grids of theoretical
model atmospheres including the computation of high resolution theoretical
spectra for stars whose physical parameters are of interest for population
synthesis. See, for example, \cite{COE03}, \cite{BER04}, \cite{MM04},
\cite{PET04}.

\begin{figure}
 \setbox20=
 \hbox{\psfig{file=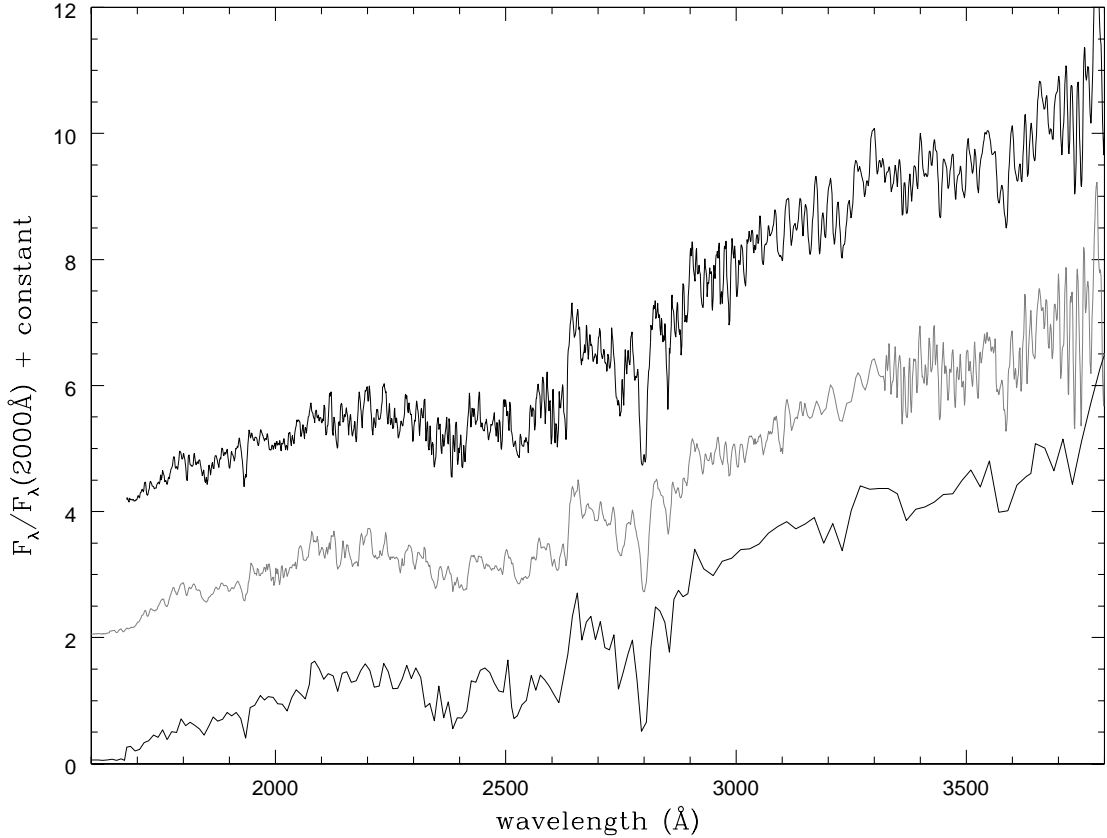,height=12cm,angle=270}}\centerline{\box20}
 \caption{BC03 standard SSP model spectra for solar metallicity at 1 Gyr in
          the wavelength range from 1600 \AA\ to 3800 \AA. The spectrum on top
	  (thick black line) is built using the HNGSL, the one in the middle
	  (gray line) uses the Pickles library and STELIB, and the bottom one
          (thin black line) the BaSel 3.1 library. The spectra have been
          normalized at 2000 \AA\ and shifted arbitrarily in the vertical
	  direction for clarity.}
\end{figure}

\begin{figure}
 \setbox20=
 \hbox{\psfig{file=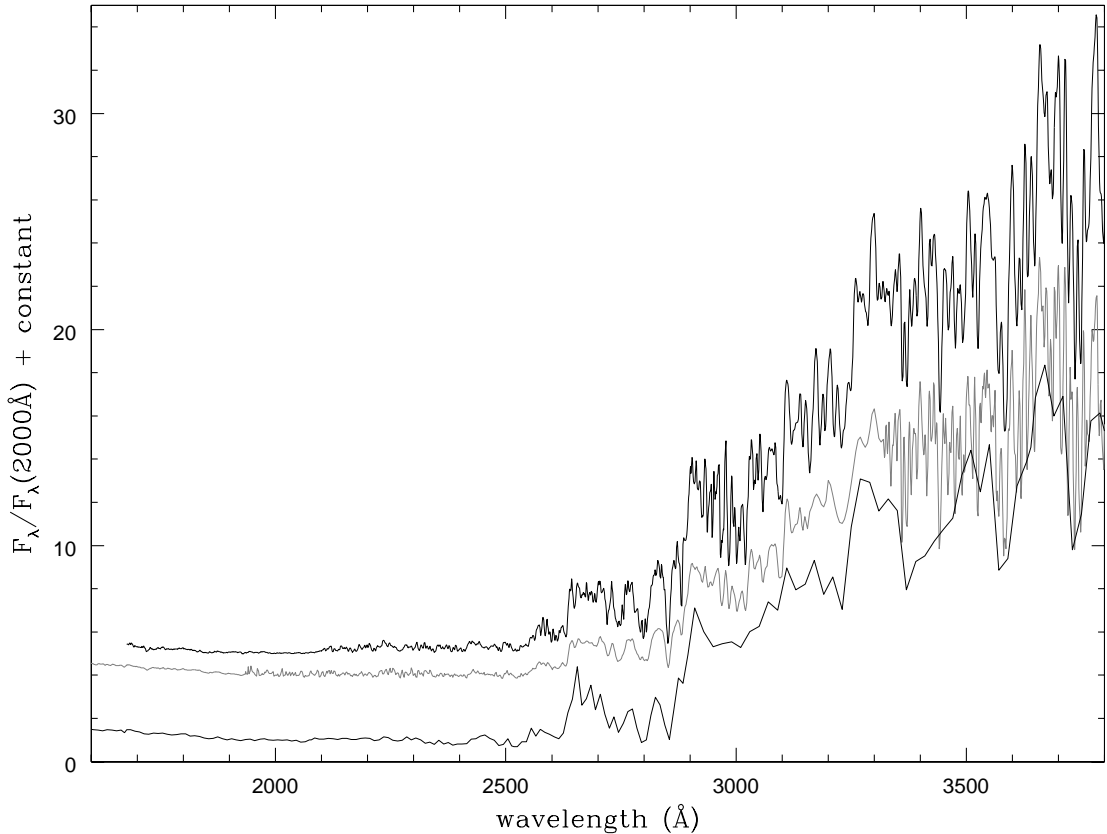,height=12cm,angle=270}}\centerline{\box20}
 \caption{BC03 standard SSP model spectra for solar metallicity at 12 Gyr in
          the wavelength range from 1600 \AA\ to 3800 \AA. The spectrum on top
	  (thick black line) is built using the HNGSL, the one in the middle
	  (gray line) uses the Pickles library and STELIB, and the bottom one
          (thin black line) the BaSel 3.1 library. The spectra have been
          normalized at 2000 \AA\ and shifted arbitrarily in the vertical
	  direction for clarity.}
\end{figure}

\section{Results}\label{sec:results}
In this section I show sample spectra from the 'standard' reference model
defined by BC03.
This model represents a simple stellar population (SSP) computed using the
Padova 1994 evolutionary tracks, the STELIB/Pickles/BaSeL 3.1 spectral libraries
and the \cite{CHAB03} IMF truncated at 0.1 and 100 M$_\odot$ (see BC03 for
details).
For illustration purposes I have computed a similar model but using the
HNGSL (\cite{HL03}) instead of STELIB to represent the stellar spectra.

Figures 1 and 2 show clearly the advantages of using the HNGSL to study
the near UV below 3300 \AA. Spectral features, lines and discontinuities,
are much better defined in the HNGSL spectra than in the IUE (used in the
Pickles library) and the BaSeL 3.1 library. However, the colors measured
from the continuum in these spectra produce very similar values. The higher
spectral resolution of STELIB above 3300 \AA\ compared to HNGSL is clearly seen 
in these figures.

In Figure 3 I compare the predicted spectra in the optical range 
for a very young (500 Myr) stellar population of solar metallicity 
using the indicated spectral libraries. Despite the very similar
continuum, the higher resolution STELIB model is to be preferred in
this optical range to study spectral lines. Figures 1 to 3 show clearly
that the BaSeL 3.1 models are not adequate at all to study spectral features,
even though they may provide a good definition of the continuum level.

The spectra in Figure 4 are useful to study the different behavior of
models computed with different libraries in the region around and above
9000 \AA. There is a clear difference in the continuum level predicted
by the HNGSL and STELIB models (lowest level) compared to the level predicted
by the BaSeL 3.1 and the Pickles library models (highest level).
The HNGSL model is to be preferred, given the higher SNR of this spectrum
in this region. However, spectral features below 8500 \AA\ are more clearly
seen in the higher resolution STELIB model.

\begin{figure}
 \setbox20=
 \hbox{\psfig{file=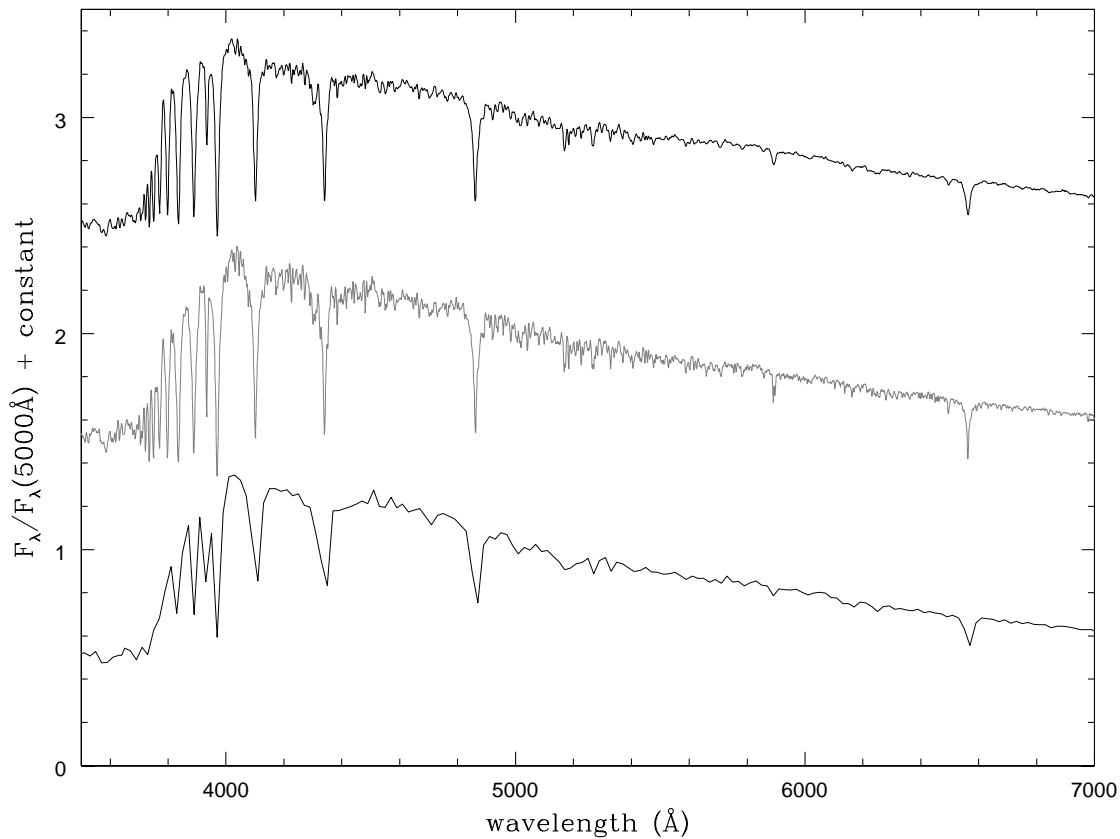,height=12cm,angle=270}}\centerline{\box20}
 \caption{BC03 standard SSP model spectra for solar metallicity at 500 Myr in
          the wavelength range from 3500 \AA\ to 7000 \AA. The spectrum on top
	  (thick black line) is built using the HNGSL, the one in the middle
	  (gray line) uses STELIB, and the bottom one
          (thin black line) the BaSel 3.1 library. The spectra have been
          normalized at 5000 \AA\ and shifted arbitrarily in the vertical
	  direction for clarity.}
\end{figure}

\begin{figure}
 \setbox20=
 \hbox{\psfig{file=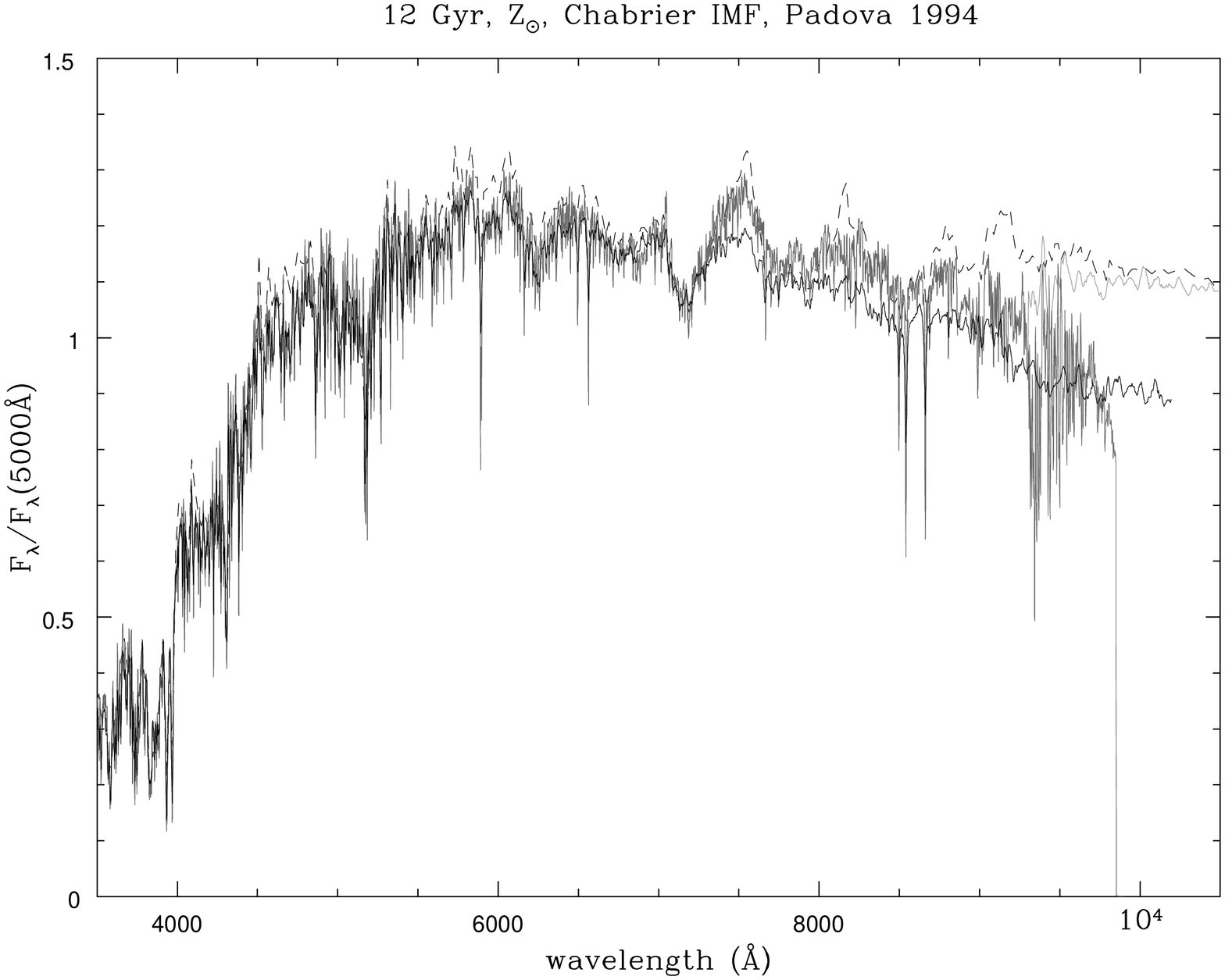,height=12cm,angle=270}}\centerline{\box20}
 \caption{BC03 standard SSP model spectra for solar metallicity at 12 Gyr in
          the wavelength range from 3500 \AA\ to 10,500 \AA.
          The black solid line (reaching up to 10,200 \AA) represents the
          spectrum built using the HNGSL,
	  the dark gray solid line (noisy above 9000 \AA) uses STELIB,
	  the light gray solid line uses STELIB up to 9000 \AA\ and the
	  Pickles library at longer wavelengths,
	  the black dashed line uses the BaSel 3.1 library.
          The spectra have been normalized at 5000 \AA.}
\end{figure}

\section{Conclusions}\label{sec:concl}
New libraries of empirical spectra of stars covering wide ranges
of values of the atmospheric parameters T$_{eff}$, log g, [Fe/H],
as well as spectral type, that have become available recently,
are complementary in spectral resolution and wavelength coverage.
Preliminary models built using the HNGSL show that this sort of
library will prove extremely useful to describe 
spectral features expected in galaxy spectra of various ages
and metallicities from the NUV to the NIR.
Due to lack of space I have compared previous models with new models
built only with the HNGSL.
Models that use all the libraries mentioned in \S2 are been built and will
be discussed in a coming paper by Bruzual \& Charlot.


\begin{thebibliography}{}

  \bibitem[Bagnulo et al. 2004] {UVES03}
     {Bagnulo, S., Jehin, E., Ledoux, C., Cabanac, R., Melo, C., Gilmozzi, R.
     and the ESO Paranal Science Operations Team} 2003,
     \comment{}
     \textit{The Messenger}, \textbf{114}, 10

  \bibitem[Bertone et al. (2004)] {BER04}
     {Bertone, E., Buzzoni, A., Rodr{\'\i}guez-Merino, L. H., \&
     Ch\'avez, M.} 2004,
     \comment{Stars at high resolution: a library of synthetic spectra from
     850 to 7000 Å}
     \textit{MmSAI}, \textbf{75}, 158

  \bibitem[Bruzual \& Charlot (2003)]{BC03}
     {Bruzual A., G. \& Charlot, S.} 2003,
     \comment{Stellar population synthesis at the resolution of 2003}
     \textit{MNRAS}, \textbf{344}, 1000

  \bibitem[Chabrier (2003)]{CHAB03}
     {Chabrier, G.} 2003,
     \comment{}
     \textit{PASP}, \textbf{115}, 763

  \bibitem[Coelho et al. (2003)] {COE03}
     {Coelho, P., Barbuy, B., Melendez, J., Allen, D. M., \& Castilho, B.} 2003,
     \comment{Compila\cc\~ao de dados at\^omicos e moleculares do UV ao IV
     pr\'oximo para uso em s{\'\i}ntese espectral}
     \textit{BASBr}, \textbf{23}, 98

  \bibitem[Heap \& Lanz 2003] {HL03}
     {Heap, S. R., \& Lanz, T.} 2003,
     \comment{HNGSL}
     \textit{in Proceedings of the ESO-USM-MPE Workshop on Multiwavelength
     Mapping of Galaxy Formation and Evolution, Venice}, ed. A. Renzini
     (in press)

  \bibitem[Heavens et al. (2004)] {HPJD04}
     {Heavens, A., Panter, B., Jim\'enez, R., \ Dunlop, J.} 2004,
     \comment{The star-formation history of the Universe from the stellar
     populations of nearby galaxies}
     \textit{Nature}, \textbf{428}, 625

  \bibitem[Le Borgne et al. (2003)] {STELIB03}
     {Le Borgne, J.-F., Bruzual A., G., Pell\'o, R., Lan{\c c}on, A.,
     Rocca-Volmerange, B., Sanahuja, B., Schaerer, D., Soubiran, C., \& 
     V{\'\i}lchez-G\'omez, R.} 2003,
     \comment{}
     \textit{A\&A}, \textbf{402}, 433

  \bibitem[Lejeune et al. (1997)] {LEJ97}
     {Lejeune, T., Cuisinier, F., \& Buser, R.} 1997,
     \comment{}
     \textit{A\&AS}, \textbf{125}, 229

  \bibitem[Lejeune et al. (1998)] {LEJ98}
     {Lejeune, T., Cuisinier, F., \& Buser, R.} 1998,
     \comment{}
     \textit{A\&AS}, \textbf{130}, 65

  \bibitem[Mateu et al. (2004)] {MAT04}
     {Mateu, J., Bruzual A., G., \& Magris, C., G.} 2004,
     \comment{}
     \textit{MNRAS} (in preparation) 

  \bibitem[Murphy \& Meiksin (2004)] {MM04}
     {Murphy, T., \& Meiksin, A.} 2004,
     \comment{A library of high resolution Kurucz spectra in the wavelength
     range 3000 - 10000 AA}
     \textit{astro-ph/0404010}

  \bibitem[Peterson et al. (2004)] {PET04}
     {Peterson, R. C., Carney, B. W., Dorman, B., Green, E. M., Landsman, W.,
     Liebert, J., O'Connell, R. W., Rood, R. T., \& Schiavon, R. P.} 2004,
     \comment{Modelling Stellar Optical and Mid-Ultraviolet Spectra from
     First Principles}
     \textit{AAS}, \textbf{204}, 07.08

  \bibitem[Pickles (1998)] {PIC98}
     {Pickles, A. J.} 1998,
     \comment{}
     \textit{PASP}, \textbf{110}, 863

  \bibitem[S\'anchez-Bl\'azquez et al. 2003] {MILES03}
     {S\'anchez-Bl\'azquez, P., Jim\'enez, J., Peletier, R., Vazdekis, A.,
     Gorgas, J., Cardiel, N., Selam, S., \& Falc\'on, J.} 2003,
     \comment{A New spectral stellar library for population synthesis}
     \textit{Rev. Mex. Astron. Astrofis. Conf. Ser.}, \textbf{17}, 192

  \bibitem[Stoughton et al. (2002)] {EDR02}
     {Stoughton et al.} 2002,
     \comment{}
     \textit{AJ}, \textbf{123}, 485

  \bibitem[Valdes et al. 2004] {VAL04}
     {Valdes, F., Gupta, R., Rose, J. A., Singh, H. P., \& Bell, D. J.} 2004,
     \comment{The Indo-US Library of Coud\'e Feed Stellar Spectra}
     \textit{ApJS}, \textbf{152}, 251

  \bibitem[Westera et al. (2002)] {WES02}
     {Westera, P., Lejeune, T., Buser, R., Cuisinier, F., \& Bruzual A., G.}
      2002,
     \comment{}
     \textit{A\&A}, \textbf{381}, 524

\end{thebibliography}
\end{document}